\documentstyle[preprint,aps]{revtex}
\tighten
\include{epsf}
\begin{document}
\preprint{Preprint Number: \parbox[t]{50mm}{ADP-98-010/T288}}
\draft
\title{Dimensionally regularized study of nonperturbative \\
         quenched QED }
\author{
   Andreas W.\ Schreiber, Tom Sizer and Anthony G.\ Williams\cite{emaddr}
}

\address{
   Special Research Centre for the Subatomic Structure of Matter \\
	 and	\\
   Department of Physics and Mathematical Physics, \\
   University of Adelaide, 5005, Australia
}
%
\maketitle
%
\begin{abstract}

  We study the dimensionally regularized fermion propagator
  Dyson-Schwinger equation in quenched nonperturbative QED in an
  arbitrary covariant gauge.  The nonperturbative fermion propagator
  is solved in $D\equiv 4-2\epsilon <4$ dimensional Euclidean space
  for a large number of values of $\epsilon$. Results for $D=4$ are
  then obtained by extrapolation to $\epsilon\to 0$.  The
  nonperturbative renormalization is performed numerically, yielding
  finite results for all renormalized quantities.  This demonstrates,
  apparently for the first time, that it is possible to
  successfully implement nonperturbative renormalization of
  Dyson-Schwinger equations within a gauge invariant regularization
  scheme such as dimensional regularization.  Here we present results
  using the Curtis-Penningon fermion-photon proper vertex for two
  values of the coupling, namely $\alpha = 0.6$ and $\alpha=1.5$ and
  compare these to previous studies employing a modified ultraviolet
  cut-off regularization.  The results using the two different
  regularizations are found to agree to within the numerical precision
  of the present calculations.
\end{abstract}

\section{Introduction}
\label{sec_intro}

Strong coupling quantum electrodynamics (QED) has been extensively studied
through the  use of the Dyson-Schwinger equations (DSE)
\cite{TheReview,MiranskReview,FGMS}.
In such studies it is unavoidable that the infinite set of coupled integral
equations be truncated to those involving Green functions with
relatively few external legs.  This truncation has 
the consequence that one or more of the Green functions appearing in the
remaining equations are no longer determined self-consistently by the
DSE's and so must be constrained, for example, by known symmetries (including 
Ward-Takahashi identities (WTI) \cite{WTI}), the absence of artificial 
kinematic singularities, the requirements of multiplicative renormalizability (MR),
and must be in agreement with perturbation theory in the weak coupling limit.
Furthermore the gauge dependence of the resulting fermion propagator
should eventually be ensured to be consistent with the Landau-Khalatnikov
transformation \cite{LKTF}.

What makes QED a particularly attractive theory to study with DSE
techniques is that the coupled integral equations determining the
photon and fermion propagators are completely closed once the
photon-fermion proper vertex is specified.  Up to transverse parts
this proper vertex is in turn determined in terms of the fermion
propagator by the corresponding WTI.  Thus, the state of the art for
this type of calculation consists of imposing the greatest possible
set of constraints and constructing the most reasonable possible
Ansatz for the transverse part of the vertex.  While such a DSE
approach to nonperturbative QED can never be an entirely
first-principles approach, such as a lattice gauge theory treatment
\cite{Rothe}, it has the advantages that it is at times possible to
obtain some analytical insights, there is no limit to the momentum
range that can be studied and one is able to compare different
regularization schemes.  This latter property will be made use of in this
paper.

A number of discussions of the choice of the transverse part of the
proper vertex can be found in the literature, e.g.,
Refs.~\cite{BC,CPI,CPII,CPIII,CPIV,dongroberts,BP1,BP2,Kiz_et_al}.  We
will concentrate here on the Curtis-Pennington (CP) vertex
\cite{CPI,CPII,CPIII,CPIV,ABGPR}, which satisfies both the WTI and the
the constraints of multiplicative renormalizability.  With a bare
vertex (which breaks both gauge invariance and MR) the critical
coupling of quenched QED differs by approximately 50\% when calculated
in the Feynman and Landau gauge. This should be compared to a
difference of less than 2\% for these gauges when calculated with the
Curtis-Pennington vertex \cite{CPIV}.  Even with the CP vertex, the
variation is significantly greater for covariant gauge choices outside
this range, particularly for negative gauges \cite{ABGPR}.  Extensions
of this work to include nonperturbative renormalization were first
performed numerically in Refs.~\cite{qed4_hw,qed4_hrw,qed4_hsw}.  In
these latter works an obvious gauge covariance violating term, arising
from the use of cut-off regularization and present in \cite{CPIV,ABGPR}
was omitted (see Refs.~\cite{dongroberts,qed4_hrw} for a discussion of
this).  Without the gauge covariance violating term the variation of the
critical coupling near the Landau gauge was again found to be rather small 
(less than 3\% when going from Landau gauge to $\xi=$0.5~\cite{qed4_hrw}).

Clearly, the gauge dependence of the (physical) critical coupling is
decreased, but not eliminated, through the use of a photon-fermion
proper vertex which satisfies the WTI. In other words, the choice of a
vertex satisfying the WTI is a necessary but not sufficient condition
in order to ensure the full gauge covariance of the Green functions of
the theory and the gauge invariance of physical observables.  The
question that arises is whether the remaining gauge dependence in the
critical coupling is primarily due to limitations of the vertex itself
or whether it is due to the use of a UV cut-off regulator in these
calculations.

Bashir and Pennington~\cite{BP1,BP2} have pursued the first of these
alternatives and have obtained, within a cut-off regularized theory,
further restrictions on the transverse part of the vertex which ensure
by construction that the critical coupling indeed becomes strictly gauge
independent.  It is not clear that adjusting the vertex to remove an
unwanted gauge-dependence is the most appropriate procedure when using
a gauge-invariance violating regularization scheme.
It should be remembered that a cutoff in (Euclidean) momentum,
apart from breaking Poincar\'e invariance, indeed also breaks
gauge invariance. It is precisely
because of this lack of gauge invariance that UV cut-off
regulators have not been used in perturbative calculations in gauge
theories for many years.  Rather, the most common perturbative method
of regularization in recent times has been that of dimensional
regularization~\cite{tHooft-Velt} where gauge invariance is explicitly
maintained.

In this work we report on a DSE study of quenched nonperturbative QED
using dimensional regularization in the renormalization procedure.
Some early exploratory studies have been carried out in the past
\cite{dim_reg}, but to our knowledge this work is the first complete
nonperturbative demonstration of dimensional regularization and
renormalization.  Nonperturbative renormalization is performed
numerically, in arbitrary covariant gauge, using the procedure first
developed and applied in Refs.~\cite{qed4_hw,qed4_hrw,qed4_hsw}.  In
these works the vertex used was that of Curtis and Pennington and so,
as it is our aim to compare to previous results obtained with the use
of cut-off regularization, we also use this vertex in the current
work. In quenched QED there is no renormalization of the
electron charge and the appropriate photon propagator is just the bare
one.  The resulting nonlinear integral equation for the fermion
propagator is solved numerically in $D = 4 - 2 \epsilon < 4$ Euclidean
dimensional space.  Successive calculations with decreasing $\epsilon$
are then extrapolated to $\epsilon=0$.

The organization of the paper is as follows:
The renormalized and dimensionally regularized SDE
formalism is discussed in Sec.~\ref{sec_formalism}.
This is followed by some representative  numerical results 
in Sec.~\ref{sec_results}. 
We present conclusions and an outlook in Sec.~\ref{sec_conclusions}.
An appendix details the final form of the fermion self-energy equations
in $D$-dimensional Euclidean space.

\section{Formalism}
\label{sec_formalism}

In this section we provide a brief summary of the implementation of
nonperturbative renormalization within the context of
numerical DSE studies.  We adopt a notation similar to that used in
Refs.~\cite{qed4_hw,qed4_hrw,qed4_hsw}, to which the reader is referred
for more detail.  The formalism is presented in
Minkowski space and the Wick rotation into Euclidean space can then
be performed once the equations to be solved have been written down.
It is important to note that although we use dimensional regularization, 
we can not make use of the popular perturbative renormalization
schemes which are usually used in connection with this, such as $MS$ or
$\overline{MS}$.  The reason, of course, is that these schemes  can only 
be defined in a purely perturbative context.

The renormalized inverse fermion propagator is defined through
\begin{eqnarray}                        \label{fermprop_formal}
            S^{-1}(\mu;p)  =   A(\mu;p^2) \not\!p - B(\mu;p^2)
            & = & Z_2(\mu,\epsilon) [\not\!p - m_0(\epsilon)]
                              - \Sigma'(\mu,\epsilon; p) \nonumber\\
            & = & \not\!p - m(\mu) - \widetilde{\Sigma}(\mu;p)\;\;\;,
\end{eqnarray}
where $\mu$ is the chosen renormalization scale, $m(\mu)$ is the value of the
renormalized mass at $p^2 = \mu^2$, $m_0(\epsilon)$ is the bare mass 
and $Z_2(\mu,\epsilon)$ is the 
wavefunction renormalization constant.  Due to the WTI for the fermion-photon
proper vertex, we have for the vertex renormalization constant
$Z_1(\mu,\epsilon)=Z_2(\mu,\epsilon)$. The renormalized and unrenormalized
fermion self-energies are denoted as $\widetilde{\Sigma}(\mu;p)$
and $\Sigma'(\mu,\epsilon;p)$ respectively.  These can be expressed in terms
of Dirac and scalar pieces, where for example
\begin{equation}
  \Sigma'(\mu,\epsilon; p) = \Sigma'_d(\mu,\epsilon; p^2) \not\!p
		     + \Sigma'_s(\mu,\epsilon; p^2)\;\;\;,
  \label{decompose}
\end{equation}
and similarly for $\widetilde{\Sigma}(\mu;p)$.
We shall for notational brevity not explicitly indicate the dependence on
$\epsilon$ of the renormalized quantities $A(\mu;p^2)$, $B(\mu;p^2)$
and $\widetilde{\Sigma}(\mu;p)$, since for these and other
renormalized quantities we will always be interested in their 
$\epsilon\to 0$ limit.  The renormalized mass function 
$M(p^2) \equiv B(\mu;p^2)/A(\mu;p^2)$ is renormalization point independent,
which follows straightforwardly from multiplicative renormalizability
\cite{qed4_hrw}. 

The renormalization point boundary condition
\begin{equation}
  \left. S^{-1}(\mu;p) \right|_{p^2 = \mu^2}
  = \not\!p - m(\mu)\:
\label{ren_point_BC}
\end{equation}
implies that $A(\mu;\mu^2) \equiv 1$ and $m(\mu) \equiv M(\mu^2)$ and yields the 
following relations between renormalized and unrenormalized self-energies
\begin{equation}\label{ren_BC}
  \widetilde{\Sigma}_{d,s}(\mu; p^2) =
    \Sigma'_{d,s}(\mu,\epsilon; p^2) - \Sigma'_{d,s}(\mu,\epsilon; \mu^2) 
     \;\;\;.
\end{equation}
Also, the wavefunction renormalization is given by
\begin{equation}
  Z_2(\mu,\epsilon) = 1 + \Sigma'_d(\mu,\epsilon; \mu^2)
\label{eq_Z2}
\end{equation}
and the bare mass $m_0(\epsilon)$ is linked to the renormalized mass 
$m(\mu)$ through
\begin{equation}
  m_0(\epsilon) = \left[ m(\mu) - \Sigma'_s(\mu,\epsilon; \mu^2) \right]
	/ Z_2(\mu,\epsilon)\;\;\;.
\label{baremass}
\end{equation}
It also follows from MR that under a renormalization point transformation
$\mu \to \mu^\prime$, $m(\mu^\prime) = M({\mu^\prime}^2)$ and
$Z_2(\mu^\prime,\epsilon) = A(\mu^\prime; \mu^2) \, Z_2(\mu,\epsilon)$
as discussed in Ref.~\cite{qed4_hrw}.

The unrenormalized self-energy is given by the integral
\begin{equation} \label{reg_Sigma}
  \Sigma'(\mu,\epsilon; p) = i Z_1(\mu,\epsilon) [e(\mu) \nu^\epsilon]^2 \int
    \frac{d^Dk}{(2\pi)^D} \gamma^\lambda {S}(\mu;k)
      {\Gamma}^\nu(\mu; k,p)
      {D}_{\lambda \nu}(\mu;p-k)\:,
\end{equation}
where $\nu$ is an arbitrary mass scale introduced in $D$
dimensions  so that the renormalized
coupling $e(\mu)$ remains dimensionless.   Since we are here working in the
quenched approximation 
we have $Z_3(\mu,\epsilon)=1$, $e_0\equiv e(\mu)$,
and the renormalized  photon propagator ${D}^{\mu \nu}(\mu;q)$ is equal
to the bare photon propagator
\begin{equation}
  D^{\mu\nu}(q) = \left (
     -g^{\mu\nu} + \frac{q^\mu q^\nu}{q^2} 
     - \xi \frac{q^\mu q^\nu}{q^2} \right )
    \frac{1}{q^2}\:
\end{equation}
with $\xi$ being the covariant gauge parameter.  Finally, 
${\Gamma}^\nu(\mu;k,p)$
is the renormalized photon-fermion vertex for which we use the CP
Ansatz, namely ($q \equiv k - p$)
\begin{equation} \label{anyfullG_eqn}
  \Gamma^\mu(\mu;k,p) = \Gamma_{\rm BC}^\mu(\mu;k,p)
    + \tau_6(\mu;k^2,p^2,q^2) \left [\gamma^{\mu}(p^2-k^2)+(p+k)^{\mu}
{\not \! q}\right ]\:,
\end{equation}
where $\Gamma_{\rm BC}$ is the usual Ball-Chiu part of the vertex 
which saturates the
Ward-Takahashi identity~\cite{BC}
\begin{eqnarray} \label{minBCvert_eqn}
  \Gamma^\mu_{\rm BC}(\mu;k,p) &=& \frac{1}{2}[A(\mu;k^2) +A(\mu;p^2)] 
\gamma^\mu
   \\
& &\hspace{1cm} + \frac{(k+p)^\mu}{k^2-p^2}
      \left\{ [A(\mu;k^2) - A(\mu;p^2)] \frac{{\not\!k}+ {\not\!p}}{2}
	      - [B(\mu;k^2) - B(\mu;p^2)] \right\}
\nonumber
\end{eqnarray}
and the coefficient function $\tau_6$ is that chosen by Curtis and Pennington,
i.e.,
\begin{equation}
  \tau_6(\mu;k^2,p^2,q^2) = -\frac{1}{2}[A(\mu;k^2) - A(\mu;p^2)] / d(k,p)\:,
\label{CPgamma1}
\end{equation}
where
\begin{equation}
  d(k,p) = \frac{(k^2 - p^2)^2 + 
[M^2(k^2)  + M^2(p^2)  ]^2}{k^2+p^2}\:.
  \label{CPgamma2}
\end{equation}

The unrenormalized scalar and Dirac self--energies are extracted out of
the DSE, Eq.~(\ref{reg_Sigma}), by taking
$\frac{1}{4}{\rm Tr}$ of this equation, multiplied by 1 and 
$\not\!p/p^2$, respectively.  Note that we use the conventions 
of Muta~\cite{Muta}

\begin{eqnarray*}
	\gamma^\mu \gamma_\mu 			&=& D
\hspace{1cm}\Rightarrow\hspace{1cm}
	\gamma^\mu \gamma^\nu \gamma_\mu	= (2-D) \gamma^\nu \\
{\rm Tr}\left [ \gamma^\mu \gamma^\nu \right ] &=& 4 g^{\mu \nu}
\hspace{2.2cm} 
{\rm Tr} [ {\bf 1} ] \> = \> 4 \\
\hspace{2.2cm}g^\mu{}_\mu\> = \>D
\end{eqnarray*}
for the Dirac algebra.

The integrands appearing in Eq.~(\ref{reg_Sigma}) only depend on the magnitude
of the internal fermion's momentum $k^2$ as well as the angle $\theta$ between
the fermion and photon momentum.  Hence the D-dimensional integrals
reduce to 2-dimensional ones, i.e.,

\begin{equation}
	\int d^D k \, f(k^2,p^2,k \cdot p) = \int d\Omega^{D-1} \times 
\int_0^\infty dk\,k^{D-1}\int_0^\pi d\theta \> \sin ^{D-2} \theta \>
 f(k^2,p^2,k \cdot p) 
\end{equation}
where 
\begin{equation}
	\int d\Omega^D = \frac{2\pi^{D/2}}{\Gamma(D/2)}
\end{equation}
is the surface area of a D-dimensional sphere.  Furthermore, it is
possible to express all the angular integrals in terms of a single
hypergeometric function so that it is only necessary to do one integral
numerically. The final form of
the regularized self-energies is presented in the appendix.

The momentum
integration is done numerically on a logarithmic grid and the renormalized
fermion DSE solved by iteration. Note that the momentum integration 
extends to infinity, necessitating a change in integration variables.  
A convenient choice of transformation is
\begin{equation}
    y = y_{\rm lo} \ \left(\frac{2}{1 - t}\right)^\frac{1}{\epsilon}\;\;\;,
\end{equation}
where $y_{\rm lo}$ is some lower integration bound and the integration variable
$t$ ranges from $-1$ to $1$. The infinite range of the
integration also requires an extrapolation of $A(\mu;p^2)$ and $B(\mu;p^2)$ 
above the highest
gridpoint.  We check insensitivity to this extrapolation by comparing results
obtained with a number of different extrapolation prescriptions.  In 
addition, we use grids which extend some 20-30 orders of magnitude beyond
what is usually used in cut-off studies.  In summary, we believe that
we have verified that
the effect of the extrapolation to infinity is well-controlled.

\section{Results}
\label{sec_results}

We present here solutions for the DSE for two values of the coupling
$\alpha = e_0^2/4\pi$, namely $\alpha=0.6$  and  $\alpha=1.5$.
These were chosen so that they correspond
to couplings respectively well below and above the critical coupling found
in previous UV cut-off based studies.
The gauge parameter is set at $\xi=0.25$, the renormalization point
which we used is $\mu^2=10^8$ and the renormalized mass is taken to be
$m(\mu) = 400$.  Note that all results are quoted in terms of dimensionless
units, i.e., all mass and momentum scales can be simultaneously multiplied
by any desired mass scale.

\begin{figure}[htb]
  \setlength{\epsfxsize}{11.5cm}
  \centering
      \epsffile{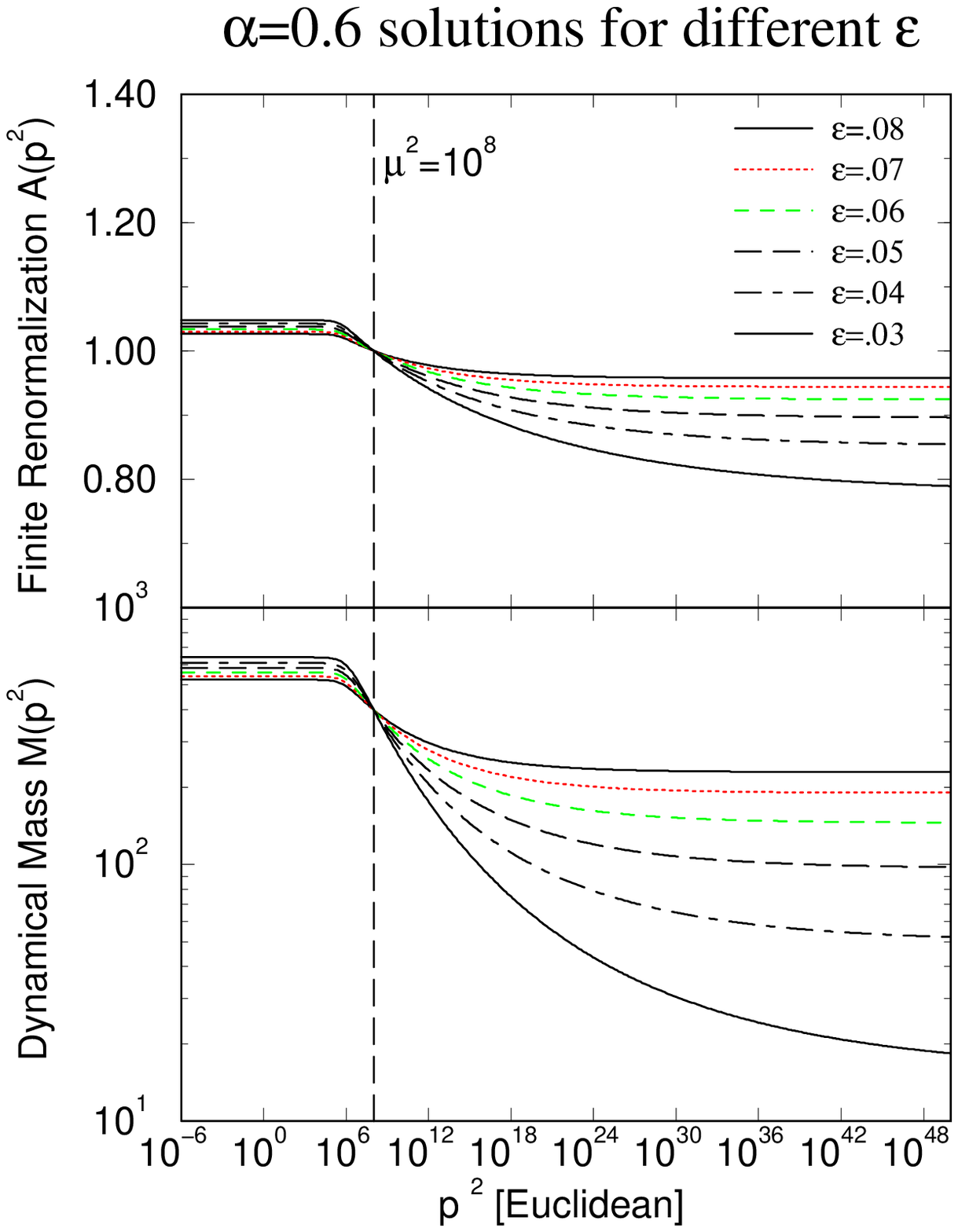}
  \parbox{130mm}{\caption{
      The finite renormalization $A(p^2)$ and mass function $M(p^2)$
      for various choices of the regulator parameter $\epsilon$.
      These results have coupling $\alpha=0.6$,
      gauge parameter $\xi=0.25$, renormalization point $\mu^2=10^8$,
      renormalized mass $m(\mu)=400$ and scale $\nu=1$. In the low $p^2$
      region the smallest $\epsilon$ has the largest value of $M(p^2)$.
  \label{sub-critical-figs}}}
  \vspace{0.5cm}
\end{figure}

\begin{figure}[htb]
  \setlength{\epsfxsize}{11.5cm}
  \centering
      \epsffile{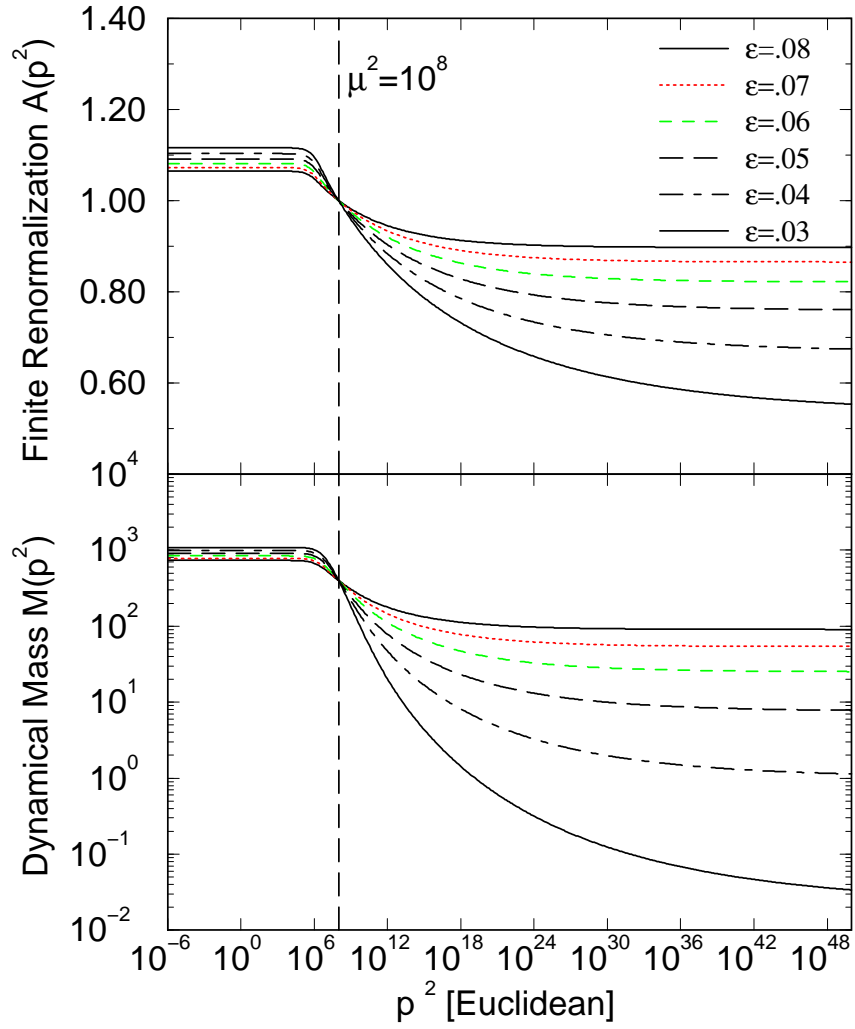}
  \parbox{130mm}{\caption{Same as Fig. 1 for $\alpha=1.5$
  \label{sup-critical-figs}}}
  \vspace{0.5cm}
\end{figure}

Figs.~\ref{sub-critical-figs} and~\ref{sup-critical-figs} show
 a family of solutions 
with the regulator parameter $\epsilon$ decreased from $0.08$ to $0.03$ for
the two values of the coupling.
 We see that the mass function increases in strength in
the infrared and tails off faster in the ultraviolet as $\epsilon$ is reduced
or $\alpha$ is increased. 

 Furthermore, it is important to note the strong
dependence on $\epsilon$, even though this parameter is already rather small.
As one would expect, the ultraviolet is most sensitive to this regulator,
however even in the infrared there is considerable dependence due to the
intrinsic coupling between these regions by the renormalization procedure.
This strong dependence on $\epsilon$ should be contrasted with the situation
in cut-off based studies where it was observed that already at rather modest
cut-offs ($\Lambda^2 \approx 10^{10}$) the renormalized functions $A$
and $M$ had reached their asymptotic limits.  At present it
is not possible to decrease
$\epsilon$ significantly below the values shown in
Figs.~\ref{sub-critical-figs} and~\ref{sup-critical-figs} because of 
limitations due to numerical noise.
In order to extract the
values of $A$ and $M$ in four dimensions we therefore need to extrapolate
to $\epsilon=0$.   More sophisticated
numerical techniques are being investigated and will hopefully allow
explicit calculations at smaller $\epsilon$ values in the future.

\begin{figure}[htb]
  \setlength{\epsfxsize}{13.0cm}
  \centering
      \epsffile{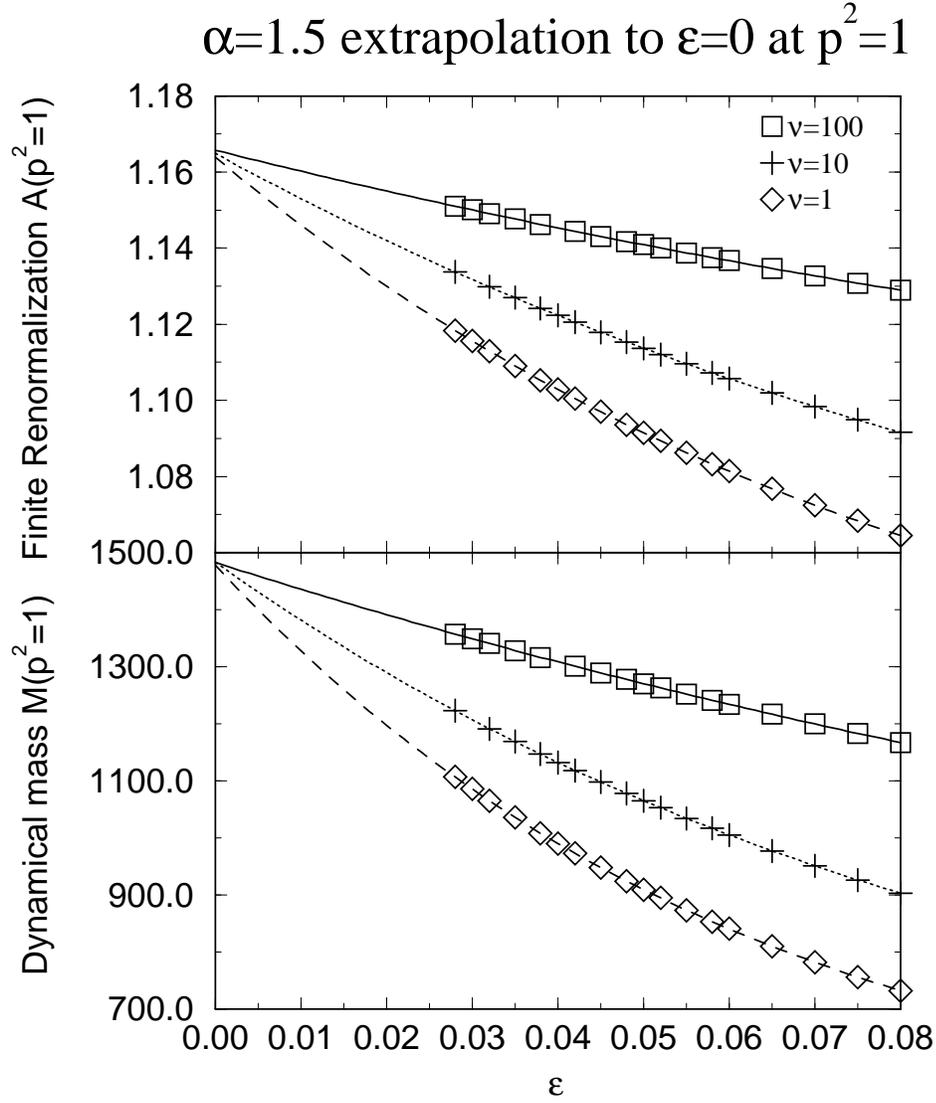}
  \parbox{130mm}{\caption{
      The finite renormalization $A(p^2)$ and mass function $M(p^2)$
      evaluated at $p^2=1$ for various values of the regulator parameter
      $\epsilon$ and extrapolated to $\epsilon=0$ by fitting a polynomial
      cubic in $\epsilon$. Shown are results for three scales $\nu=1$, $10$
      and $100$. All other parameters are those of Fig. 2. The different scales
      coincide at $\epsilon=0$ to an accuracy of approximately  $0.2\%$.
  \label{sup-critical-ir-fit}}}
  \vspace{0.5cm}
\end{figure}

An extrapolation such as this always involves an added uncertainty in
the final result.  It is fortunate that it is possible to
estimate this uncertainty by making use of the fact that in the limit
$\epsilon \rightarrow 0$ the renormalized quantities should become
independent of the arbitrary scale $\nu$, which was 
introduced to keep the coupling $\alpha$ dimensionless in $D$ dimensions. In
Fig.~\ref{sup-critical-ir-fit} we show $A(p^2)$ and $M(p^2)$ evaluated
with $\alpha=e_0^2/4\pi=1.5$ in the infrared (at $p^2=1$) as a function of
$\epsilon$ for a range of values
of $\nu$. The results at $\epsilon=0$ are extracted
from cubic polynomial fits in $\epsilon$.  As may be observed, the
agreement between the different curves at $\epsilon=0$ is excellent,
being of the order of $0.2\%$.

\begin{figure}[htb]
  \setlength{\epsfxsize}{13.0cm}
  \centering
      \epsffile{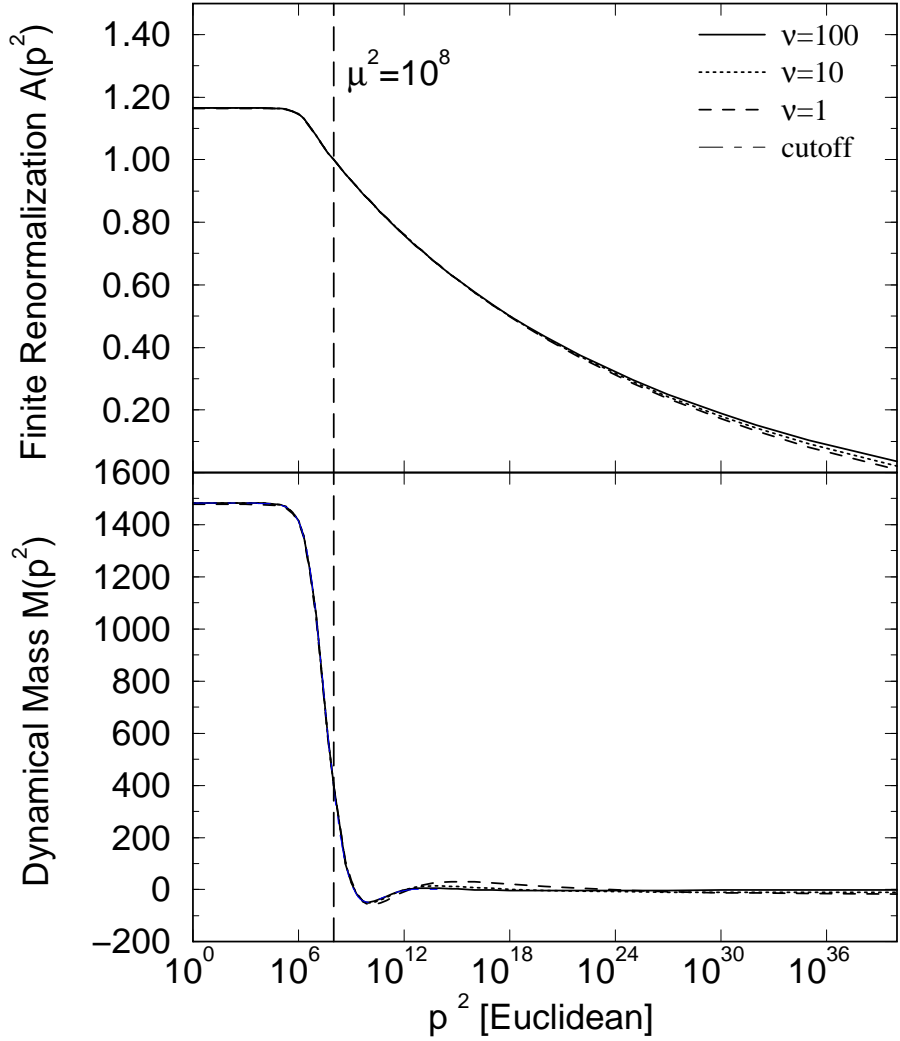}
  \parbox{130mm}{\caption{
      The finite renormalization $A(p^2)$ and mass function $M(p^2)$
      extrapolated to $\epsilon=0$ at every momentum point for three
      different scales $\nu=1$, $10$ and $100$.
      They were calculated by fitting a  cubic polynomial in $\epsilon$
      at each momentum point.
      All other parameters are those of Fig. 2. Also shown is the
      result obtained using a modified UV cutoff. The UV cutoff and 
      extrapolated $\nu=100$ results are indistinguishable in these
      plots.
  \label{sup-critical-fit}}}
  \vspace{0.5cm}
\end{figure}

In Fig.~\ref{sup-critical-fit} we show the results extrapolated to
$\epsilon=0$ as a function of the momentum (again for $\alpha = 1.5$).
Also shown, although hardly distinguishable, is the result for these
curves as obtained in the modified UV cutoff based studies, which used a
gauge-covariance fix to remove an obvious part of the gauge dependence
induced by the cutoff \cite{qed4_hrw}. Again the agreement is very
good for a wide range in $p^2$ and $\nu$.  Only in the ultraviolet
region (above say $p^2 = 10^{12}$) do differences between the curves
become discernible.  We see in Fig.~\ref{sup-critical-fit} that the UV
cutoff result is almost indistinguishable from the $\nu=100$ result at
essentially all momenta. The discrepancies for the $\nu=1$ and $10$
cases are greater in the UV due to the fact that in this region the
errors introduced by the extrapolation procedure in $\epsilon$ become
comparable (for a cubic fit) to the functions' values there.  This is
supported by the following two observations. Firstly, from
Fig.~\ref{sup-critical-ir-fit} we see that the $\epsilon$-dependence
is almost linear for $\nu=100$, whereas it clearly deviates from
linearity for the other cases.  Secondly, there is little change in
the $\nu=100$ extrapolation as we change the order of the fit
polynomial up to order $\epsilon^6$, whereas the $\nu=1$ and $10$
results show greater variation.  We conclude that the
$\nu=100$ results in Fig.~\ref{sup-critical-ir-fit} provide the
most reliable $\epsilon\to 0$ extrapolation.  It should also
be noted that the oscillatory behaviour in the mass function first
noticed in~\cite{qed4_hw} is reproduced in this work, and
the results for the unmodified UV cutoff disagree with those from
dimensional regularization.

Finally, we present in Tables \ref{sub_tbl} and \ref{sup_tbl} the bare
mass $m_0(\epsilon)$ and the wavefunction renormalization for the two
cases studied as a function of the regularization parameter $\epsilon$
and the gauge parameter $\xi$.  It should be noted that these two
quantities are by their very nature sensitive to the behaviour of
$A(p^2)$ and $B(p^2)$ in the ultraviolet.  In particular, while any
renormalized quantities were found to have a negligible
dependence on the precise form of the extrapolation of the integrands
beyond the highest gridpoint, it was found that $Z_2(\mu,\epsilon)$
and $m_0(\epsilon)$ do show some small dependence ($<1\%$) in our
calculations.  The values for these quantities listed in Tables
\ref{sub_tbl} and \ref{sup_tbl} were obtained by assuming a simple
power-behaviour of the integrands beyond the highest gridpoint.  As
was seen in the UV cutoff studies, the wavefunction renormalization
(for $\xi > 0$) actually decreases as the regularization is removed.
In the same way the behaviour of $Z_2$ as a function of the gauge
parameter is qualitatively the same as observed in the cut-off
studies, i.e. at fixed $\epsilon$ it decreases as one moves from the
Landau towards the Feynman gauge.  The bare mass $m_0$, on the other
hand, appears to show a different behaviour to before, at least for
large couplings: while in the present work it decreases as the gauge
parameter is increased for all $\epsilon$, in cut-off studies it only
did this for moderately small cut-offs.  It is quite possible that
this different qualitative behaviour reflects the fact the even the
lowest $\epsilon$'s which have been reached here ``correspond'' to
rather modest cut-offs.  In connection with this note that at the
values of $\epsilon$ shown in Tables \ref{sub_tbl} and \ref{sup_tbl}
the mass functions in Figs.~\ref{sub-critical-figs}
and~\ref{sup-critical-figs} are still positive everywhere, the
oscillations only setting in as one extrapolates toward $\epsilon
\rightarrow 0$.

\begin{table}
  \caption{ Renormalization constant $Z_2(\mu,\epsilon)$ and bare mass
    $m_0(\epsilon)$ as functions of regularization parameter $\epsilon$
    for $\alpha = 0.6$ in various gauges $\xi$.  All solutions are with
    renormalization point $\mu^2 = 1.00\times 10^{8}$ and renormalized mass
    $m(\mu) = 400.0$}
  \setdec 0.0000
  \begin{tabular}{r@{.}l@{\hspace{10mm}}|r@{.}lr@{.}lr@{.}l@{\hspace{10mm}}|r@{.}l@{$\times$}lr@{.}l@{$\times$}lr@{.}l@{$\times$}l}
    \multicolumn{2}{c|}{$\epsilon$}          &
    \multicolumn{6}{c|}{\hspace{-1.0cm}$Z_2(\mu,\epsilon)$} &
    \multicolumn{9}{c}{$m_0(\epsilon)$}      \\
  \tableline
    \multicolumn{2}{c|}{}           &
    \multicolumn{2}{c} {$\xi=0$   } &
    \multicolumn{2}{c} {\hspace{-0.25cm}$\xi=0.25$} &
    \multicolumn{2}{c|}{\hspace{-1.0cm}$\xi=0.5$ } &
    \multicolumn{3}{c} {$\xi=0$   } &
    \multicolumn{3}{c} {$\xi=0.25$} &
    \multicolumn{3}{c} {$\xi=0.5$ } \\
  \tableline
   0&08 & 1&000 & 0&957 & 0&917 & 2&30 & $10^{2}$ & 2&29 & $10^{2}$ & 2&28 & $10^{2}$ \\
   0&07 & 1&000 & 0&944 & 0&891 & 1&91 & $10^{2}$ & 1&90 & $10^{2}$ & 1&89 & $10^{2}$ \\
   0&06 & 1&000 & 0&924 & 0&854 & 1&46 & $10^{2}$ & 1&45 & $10^{2}$ & 1&44 & $10^{2}$ \\
   0&05 & 1&000 & 0&896 & 0&802 & 9&75 & $10^{1}$ & 9&66 & $10^{1}$ & 9&57 & $10^{1}$ \\
   0&04 & 1&000 & 0&851 & 0&725 & 5&07 & $10^{1}$ & 5&00 & $10^{1}$ & 4&94 & $10^{1}$ \\
   0&03 & 1&000 & 0&779 & 0&606 & 1&59 & $10^{1}$ & 1&56 & $10^{1}$ & 1&53 & $10^{1}$ \\
  \end{tabular}
  \label{sub_tbl}
\end{table}

\begin{table}
    \caption{ As above, but for $\alpha = 1.5$}
  \begin{tabular}{r@{.}l@{\hspace{10mm}}|r@{.}lr@{.}lr@{.}l@{\hspace{10mm}}|r@{.}l@{$\times$}lr@{.}l@{$\times$}lr@{.}l@{$\times$}l}
    \multicolumn{2}{c|}{$\epsilon$}          &
    \multicolumn{6}{c|}{\hspace{-1.0cm}$Z_2(\mu,\epsilon)$} &
    \multicolumn{9}{c}{$m_0(\epsilon)$}      \\
  \tableline
    \multicolumn{2}{c|}{}           &
    \multicolumn{2}{c} {$\xi=0$   } &
    \multicolumn{2}{c} {\hspace{-0.25cm}$\xi=0.25$} &
    \multicolumn{2}{c|}{\hspace{-1.0cm}$\xi=0.5$ } &
    \multicolumn{3}{c} {$\xi=0$   } &
    \multicolumn{3}{c} {$\xi=0.25$} &
    \multicolumn{3}{c} {$\xi=0.5$ } \\
  \tableline
   0&08 & 1&000 & 0&897 & 0&805 & 9&20 & $10^{1} $ & 9&06 & $10^{1} $ & 8&92 & $10^{1} $ \\
   0&07 & 1&000 & 0&865 & 0&749 & 5&51 & $10^{1} $ & 5&40 & $10^{1} $ & 5&28 & $10^{1} $ \\
   0&06 & 1&000 & 0&821 & 0&674 & 2&58 & $10^{1} $ & 2&51 & $10^{1} $ & 2&44 & $10^{1} $ \\
   0&05 & 1&000 & 0&759 & 0&576 & 7&94 & $10^{0} $ & 7&60 & $10^{0} $ & 7&28 & $10^{0} $ \\
   0&04 & 1&000 & 0&669 & 0&447 & 1&10 & $10^{0} $ & 1&02 & $10^{0} $ & 9&54 & $10^{-1}$ \\
   0&03 & 1&000 & 0&535 & 0&286 & 2&63 & $10^{-2}$ & 2&27 & $10^{-2}$ & 1&96 & $10^{-2}$ \\
  \end{tabular}
  \label{sup_tbl}
\end{table}

\section{Conclusions and Outlook}
\label{sec_conclusions}

We have reported here the first detailed study of the numerical
renormalization of the fermion Dyson-Schwinger equation of QED
through the use of a dimensional regulator rather than a
gauge invariance-violating UV cut-off.  The initial results
presented here are encouraging.  Firstly, we have explicitly demonstrated
that the approach works and is independent of the intermediate
dimensional regularization scale ($\nu$) as expected.  Secondly, we
have seen the interesting result that our calculations using dimensional
regularization agree with modified UV cut-off calculations within
the current numerical precision, but disagree with the unmodified
UV cutoff ones.

A significant practical difference between the dimensionally and UV
cut-off regularized approaches is that in the former it is at present
necessary to perform an explicit extrapolation to $\epsilon=0$ whereas
in the latter it was found that for a sufficiently large choice of UV
cut-off the results became independent of the cut-off.  This need to
extrapolate, together with the high precision that one needs to attain
in order to make meaningful comparisons with cut-off based studies,
makes numerical dimensional regularization and renormalization of
Schwinger-Dyson equations a rather formidable task.  Nevertheless,
although we are presently investigating whether it is numerically
possible to extend the studies to even smaller values of $\epsilon$ in
order to improve the precision of the $\epsilon\to 0$ extrapolation,
the extrapolation to $\epsilon=0$ appears to be well under control, at
least for values of the fermion momentum away from the ultraviolet
region.

Having demonstrated the numerical procedure of renormalization
using dimensional regularization, we now plan to study chiral symmetry
breaking and in particular hope to extract the critical coupling
as a function of the gauge parameter.  Results of this ongoing
work will be presented elsewhere.  The eventual aim is to extend this
treatment to the case of unquenched QED and to a systematic study
of electron-photon proper vertices.

\begin{acknowledgements}

This work was partially supported by grants from the Australian Research 
Council and by an Australian Research Fellowship.

\end{acknowledgements}

\begin{appendix}

\section{Final form for the regularized fermion self-energies}
\label{appdx_a}

In the quenched approximation, all angular integrals in the Dirac and scalar
regularized self-energies defined by the Euclidean analogue of
Eq.~(\ref{reg_Sigma})  may be expressed in terms of  the integrals
\begin{equation}
	I_n^D(w) = {\Gamma \left ({D \over 2} \right )
\over \Gamma \left ({D-1 \over 2} \right ) \sqrt \pi}
\int_0^\pi d\theta\,  \left ({x+y \over z}\right )^n \, \sin^{D-2} \theta \, 
    \quad\text{for }n=-1,1,2
\end{equation}
where we have defined dimensionless quantities $x = p^2/\nu^2$, $y = k^2/\nu^2$, $z = q^2/\nu^2 = x+y-2 \sqrt{x y}
\cos \theta$  and $w$=$y/x$. Similarily, we shall for convenience
define dimensionless
versions of $A(\mu^2;p^2)$ and $B(\mu^2;p^2)$, namely $a(x)=A(\mu^2;p^2)$ and 
$ \nu b(x) = B(\mu^2;p^2)$ (we suppress the dependence on $\mu$ here in 
order to  make the notation less cumbersome).
Explicit evaluation (for $0 \leq w \leq 1$) yields 
\begin{eqnarray}
I_{-1}^D(w) & = & 1 \\ \nonumber
I_{1}^D(w) & = & (1+w) \,\,{}_2F_1(1,\epsilon;2-\epsilon;w)\\ \nonumber
I_{2}^D(w) & = & 2 \left ({1+w \over 1-w}\right )^2 \left (
 -{1 - 2 \epsilon  \over 2}  I^D_1(w) + 1 - \epsilon\right ) \,,
\end{eqnarray}
and for $w \geq 1$ one may use
\begin{equation}
	I_n^D(w) = I_n^D (w^{-1} )\;\;\;.
\end{equation}
Defining 
\begin{eqnarray}
	\widetilde{d_E}(x,y) &=& \frac{d_E(x,y)}{x+y}			\\
	\widetilde{\Delta{a}}(x,y) &= &
		\frac{a(x) - a(y)}{(1-w)/(1+w)} 
\end{eqnarray}
(and similarly for $\widetilde{\Delta{b}}(x,y)$)
the regularized self-energies in the quenched
approximation become
\begin{eqnarray}
  \Sigma'_d(\mu,\epsilon; x)
  & = & 
 	\frac{\alpha_0}{4\pi} (4\pi)^\epsilon
	\frac{Z_1(\mu,\epsilon)}{\Gamma(2-\epsilon)}
	\int_0^\infty dy \frac{y^{-\epsilon}}{y a^2(y) + b^2(y)}\,w
							\nonumber \\
  & \times &
      \Bigg\{
	a(y) \frac{a(x) + a(y)}{2} \xi
	        \left(1-\epsilon\right) \left(1 - I_1^D(w)
		\right) 				\nonumber \\
  & &	+ a(y) \frac{\widetilde{\Delta a}(x,y)}{2}
		\left[ \Bigl(
		       {1 \over 2} + \left(1-\xi\right) \left(1-\epsilon\right)
		       - {3-2 \epsilon \over 2 \tilde d_E(x,y) }
				\left(\frac{1-w}{1+w}\right)^2
		       \Bigr)
		       \left(I_1^D(w) - 1\right) \right.\nonumber \\
  && \left.\hspace{5cm}
		      + \frac{2 \xi w}{(1+w)^2} I_1^D(w)
		\right]					\nonumber \\
  & &	+ \frac{b(y)}{x+y} \widetilde{\Delta b}(x,y)
		\left[ \Bigl(
		       {1 \over 2}
		       + \left(1-\xi\right) \left(1-\epsilon\right)
		       \Bigr)
		       \left(I_1^D(w) - 1\right)
		      + \frac{\xi}{1+w} I_1^D(w)
		\right]
      \Bigg\}					    \label{DR_red_sigpd}
\end{eqnarray}
and
\begin{eqnarray}
  \Sigma'_s(\mu,\epsilon; x)
  & = & 
	\frac{\alpha_0}{4\pi} (4\pi)^\epsilon
	\frac{Z_1(\mu,\epsilon)}{\Gamma(2-\epsilon)}
	\int_0^\infty dy \frac{y^{1-\epsilon}}{y a^2(y) + b^2(y)}
	\frac{1}{x(1+w)}				\nonumber \\
  & \times &
      \Bigg\{
	b(y) \frac{a(x) + a(y)}{2}
	        \left(3 - 2 \epsilon +\xi\right) I_1^D(w)	\nonumber \\
  & &	- a(y) \widetilde{\Delta b}(x,y)
		\left[ \Bigl(
			{1 \over 2} + \left(1-\xi\right) \left(1-\epsilon\right)
		       \Bigr) 
		       \left(I_1^D(w) - 1\right)
		      + \frac{\xi w}{1+w} I_1^D(w)
		\right]					\nonumber \\
  & &	+ b(y) \widetilde{\Delta a}(x,y)
		\left[ \Bigl(
		       {1 \over 2} + \left(1-\xi\right) \left(1-\epsilon\right)
		       \Bigr)
		       \left(I_1^D(w) - 1\right)\right.\nonumber \\
  && \left.\hspace{5cm}
		      + \left( \frac{\xi}{2} +
			    \frac{3-2 \epsilon}{2\widetilde{d_E}(x,y)}
				\left(\frac{1-w}{1+w}\right)^2
			\right)I_1^D(w)
		\right]
      \Bigg\}					    \label{DR_red_sigps}
\end{eqnarray}

It should be remembered that the equivalent expressions in cut-off regularized
quenched QED suffer from an ambiguity due to the lack of gauge invariance
in those calculations: one obtains different results at this stage
depending one whether or not one has made use of the Ward-Takahashi identity
in the initial stages of the calculation\cite{dongroberts,BP1}.
The difference shows up in the terms multiplied by $a^2(y)$ and $b^2(y)$.
Readers may readily convince themselves that no such ambiguity exists
in the present work.  In order to do this, the following identity
is of use:
\begin{equation}
0 \> = \> \int_0^\infty dy \, y^{1 - \epsilon} 
\left ( 2 {1 - \epsilon \over y-x} \> + \> 
\left [ {2 \epsilon - 1 \over y-x} - {1 \over x+y} \right ] 
I_1^D(w) \right )
\end{equation}
\end{appendix}


\newpage


\end{document}